\documentclass[12pt]{article}
\usepackage{amsfonts}

\newcommand{\cl}{{\cal{L}}}

\usepackage{amsmath}
\usepackage{amssymb}
\newcommand{\be}{\begin{equation}}
\newcommand{\ee}{\end{equation}}
\newcommand{\bea}{\begin{eqnarray}}
\newcommand{\eea}{\end{eqnarray}}
\newcommand{\ba}{\begin{array}}
\newcommand{\ea}{\end{array}}

\newcommand{\wt}{\widetilde}

\title {Leptgenesis from Bilinear R-parity Violating Couplings}
\author {Ido Ben-Dayan,\\
 Department of Particle Physics,  Weizmann Institute of Science,\\
Rehovot 76100, Israel.\\
\begin{small}email:ido.ben-dayan@weizmann.ac.il\end{small}}

\begin{document}\maketitle



\vspace{5ex}
{\begin{center}
{\huge\textbf{Abstract}\\}
\end{center}}
\vspace{5ex} \small We reexamine the idea that bilinear R-parity
violating couplings can be responsible for leptogenesis. We prove that, to
have lepton number violation before the electroweak phase transition,
misalignment between the lepton-Higgsino and slepton-Higgs sectors
must be involved.  The processes that generate a lepton asymmetry are
bino or wino decays into a lepton-Higgs and slepton-Higgsino final
states. Since these decays occur mainly while the gauginos are still
in thermal equilibrium, they quantitatively fail to produce the
observed baryon asymmetry of the universe.

\section{Introduction}
The observed baryon asymmetry is \cite{PDG:2004}: \be B \equiv
\frac{n_b - n_{\bar{b}}}{s}= \frac{n_b}{s}\simeq 10^{-10}. \ee It is
possible that the origin of this asymmetry is a lepton asymmetry,
which is partially converted into a baryon asymmetry by sphaleron
processes \cite{FY:1986}. This scenario is known as leptogenesis.
Within the framework of the supersymmetric standard model (SSM), the
required lepton number violation can be induced by bilinear R-parity
violating couplings \cite{Ma:2000,Hambye:2000zs,Ma:2000wp}. We
critically examine this idea, paying special attention to the
condition that the lepton number violating decays must occur out of
equilibrium \cite{Sakharov:1967}.

Within the SSM, the final baryon asymmetry is related to the initial
lepton asymmetry by $B =-\frac{32}{92}L$
\cite{Harvey:1990,Dreiner:1993}. The lepton asymmetry can be
expressed as a product of three factors: the ratio between the
number density of the decaying particle $n_i$ and the entropy
density $g_* n_\gamma$ ($g_*$ denotes the effective number of
relativistic degrees of freedom), the branching ratio into lepton
number violating modes ${\cal B}^{\not L}_i$, and the CP asymmetry
in these decays $\epsilon_i$:
 \be  L \simeq {\cal B}^{\not L}_i\times \epsilon_i\times
 \frac{n_i}{g_*n_{\gamma}}.
 \ee
We will estimate each of these three factors in order to obtain an
estimate or, more precisely, an upper bound, on the lepton asymmetry
so induced. The plan of this paper is as follows. In Section 2 we
define the model, and find the sources of CP and L violation. In
Section 3 we estimate the branching ratio ${\cal B}^{\not L}_i$. In
Section 4 we find the CP asymmetry $\epsilon_i$. In Section 5 we
obtain an upper bound on $n_i$ at the time of departure from thermal
equilibrium. In section 6 we obtain an upper bound on the lepton
asymmetry generated in this model and conclude.

The idea of leptogenesis from bilinear R-parity violating couplings
was originally proposed and investigated in
refs. \cite{Ma:2000,Hambye:2000zs,Ma:2000wp}. We disagree with their
results in several points and, in particular, the final answer on
whether this could be a successful scenario for baryogenesis.

\section{CP and Lepton Number Violation}
We work in the framework of the SSM without R-parity
[$R_p=(-1)^{3(B-L)+2s}$]. We follow the notation and methods of refs
\cite{Martin:1999,Banks:1995}. The SSM has four chiral
supermultiplets that are doublets of $SU(2)$ and carry hypercharge
$-1/2$. These are the ``down Higgs" $H_d$ and the ``lepton-doublet"
$L_i$  superfields. In the absence of R-parity, there is no quantum
number to distinguish the Higgs from the leptons, so we denote these
fields collectively as $L_{\alpha}$, with $\alpha=0,1,2,3$. We
neglect trilinear R-parity violating terms in the superpotential and
in the soft SUSY breaking part of the Lagrangian. The relevant part
of our lagrangian is the following:
\begin{eqnarray}
\cl &\supset& \left(\imath\frac{g_1}{\sqrt{2}}\tilde
B+\imath\sqrt{2}g_2\tilde W^aT^a\right)\left(H_u^{\dag}\tilde
H_u+L^{\dag}_{\alpha}\tilde L_{\alpha}\right)
-\frac{1}{2}\mu_{\alpha}\tilde L_{\alpha}\tilde
H_u\nonumber\\
&-&\left(|\mu_{\alpha}|^2+m_{H_u}^2\right)|H_u|^2-\left(m_{\alpha
\beta}^2+\mu_{\alpha} \mu_{\beta}^{\dag}
\right)L_{\alpha}L_{\beta}^{\dag}-b_{\alpha}L_{\alpha}H_u\nonumber\\
&-&\frac{1}{8}(g_1^2+g_2^2)\left(|H_u|^2-|L_{\alpha}|^2
\right)^2-\frac{1}{2}m_{\tilde B}\tilde B \tilde B
-\frac{1}{2}m_{\tilde W}\tilde W^a \tilde W^a+h.c.
\end{eqnarray}
where $\tilde X(X)$ denotes fermions (scalars). Without loss of
generality, we can
choose a basis in which $\mu_{\alpha}=\mu(1,0,0,0)$ and $m^2_{I \neq J}=0$
where $I,J=1,2,3$. With additional phase rotations to remove
unphysical phases, we finally have the following lagrangian:
\begin{eqnarray}
\cl &\supset& [\imath\frac{g_{1}}{\sqrt{2}} L^{\dag}_{\alpha}\tilde
L_{\alpha} \tilde B+\imath\frac{g_{1}}{\sqrt{2}} H^{\dag}_u\tilde
H_u\tilde B+\imath\sqrt{2}g_2e^{i \phi_g}
L^{\dag}_{\alpha} \tilde L_{\alpha}T^a\tilde
W^a
+\imath\sqrt{2}g_2e^{i \phi_g}H^{\dag}_u\tilde H_uT^a\tilde W^a \nonumber \\
&-&|b_{\alpha}|L_{\alpha}H_u
 -\frac{1}{2}|\mu|e^{i\phi_0}\tilde L_0\tilde H_u-|m_{0I}^2|e^{i\phi_I}L_0^{\dag}L_I \nonumber \\
 &-&\frac{1}{2}|m_{\tilde B}|\tilde B\tilde B
 -\frac{1}{2}|m_{\tilde W}|\tilde W^a\tilde W^a+h.c]+\mbox{terms with
real couplings.}
\end{eqnarray}
The five physical phases are
$ \phi_g =  \frac{1}{2}arg\left(\frac{m_{\tilde B}}{m_{\tilde W}}\right),
 \phi_0 =  arg \left(\frac{\mu m_{\tilde B}}{b_0}\right)$ and
 $\phi_I= arg \left(\frac{m^2_{0I}b_0}{b_I}\right)$.
The CP violating phase that would turn out to induce leptogenesis is
$\phi_g$.

We now identify the conditions for lepton number violation. Since
R-parity is violated here only by bilinear terms, lepton number
violation will be induced by the mass matrices. Note that at
temperatures above the electroweak phase transition (EWPT), the SU(2)
symmetry is unbroken ($\langle H_u\rangle=\langle L_\alpha\rangle=0$).
Different SU(2) multiplets do not mix. Thus the question of whether
lepton number is violated can be examined by considering the SU(2)
doublets alone. The fermion mass matrix in the basis $(\tilde
H_u,\tilde L_\alpha)$ has the form
 \\
$M^f= \left(%
\begin{array}{ccc}
 0 & \mu & 0_{1 \times 3}\\
 \mu & 0 & 0_{1 \times 3}\\
 0_{3 \times 1} & 0_{3 \times 1}
  & 0_{3 \times 3}
\end{array}%
\right)$
\\
Thus, defining $\tilde L_0$ (the field that has a $\mu$-term that
couples it to $H_u$) as the Higgs field $\tilde H_d$, and the three
$\tilde L_I$ fields (the ones orthogonal to the direction of $\mu$)
as leptons, we learn that lepton number is conserved in the fermion
sector. In other words, the leptons are massless above the EWPT
while the Higgsinos are massive.

Consider now the mass-squared matrix of the $(H_u,L_\alpha)$ scalars:
\\
 $M^s=
 \left(\begin{array}{ccc}
  |\mu|^2+m_{H_u}^2 & b_0 & b_I \\
  b_0 & |\mu|^2+m_{H_d}^2 & m_{I0}^2 \\
   b_I^{\dag} & m_{I0}^{2\dag} & m_{L_I}^2 \\
 \end{array}\right)$
\\
The lepton number, defined above, is violated by the $b_I$ and
$m^2_{I0}$ terms.  We conclude that, for lepton number to be
violated before the EWPT, one must consider both the scalar and
fermionic sectors of the model. The basis invariant statement is
that, in the SU(2)-doublet sector, a misalignment between the
fermion and the scalar mass matrices is the source of lepton number
violation. In \cite{Ma:2000,Hambye:2000zs,Ma:2000wp} the authors
neglected the mixing in the bosonic sector which leads to a
conserved lepton number.

\section{The Branching Ratio ${\cal B}^{\not L}$}
To calculate the branching ratio of lepton number violating gaugino
decays, we switch to the mass basis:
 \be
    \begin{pmatrix}
      H_u'\\
      H_d'\\
      L_I'\\
    \end{pmatrix}=U^s \times \begin{pmatrix}
     H_u\\
      H_d\\
      L_I\\
    \end{pmatrix},\ \ \ \ \ \
    \begin{pmatrix}
    \tilde  H'_u\\
     \tilde H_d'\\
      \tilde L_I'\\
    \end{pmatrix} = U^f \times \begin{pmatrix}
    \tilde H_u\\
     \tilde H_d\\
      \tilde L_I\\
    \end{pmatrix}.
    \ee
Here $U^f$ and $U^s$ are the unitary matrices which diagonalize $M^f$
and $M^s$, respectively, and primes denote mass eigenstates. In this
basis, the gaugino-matter couplings can be written as follows:
\be
  \imath \sqrt{2}g_{(k)}T^a \tilde
\lambda^a_{(k)} A^{\dag}_l \delta_{lk} \psi_k = \imath
\sqrt{2}g_{(k)}T^a \tilde \lambda^{a}_{(k)} A^{'\dag}_m
V_{mn}\psi'_n \ee where $\tilde \lambda^{a}$ denote gauginos, $A'$
and $\psi'$ denote, respectively, scalar and fermion matter fields.
Before the EWPT, the gauginos do not mix, so that $\tilde
\lambda^{a}$ stands for both the interaction and mass eigenstates.
The mixing matrix $V=U^sU^{f\dag}$ is unitary. The important point
is that if the $M^s$ and $M^f$ matrices were aligned, namely $b_I$
and $m^2_{0I}$ had vanished in the basis where $\mu_I=0$, then $U^s$
and $U^f$ would be both block diagonal, with
$(U^s)_{LH}=(U^f)_{LH}=0$ (by $L$ we mean here any of the three
$L_I$ and by $H$ we refer to $H_u$ and $H_d$), and so would $V$.
With misaligned matrices, we have lepton number violating elements,
$V_{LH}\neq0$. These lead to gaugino decays which violate lepton
number: \be {\cal B}^{\not
  L}\equiv BR(\tilde \lambda\rightarrow LH) \simeq
\frac{\sum_{L,H}|V_{LH}|^2}{5}\ee
where $\tilde\lambda=\tilde B$ or $\tilde W$ and, if the masses of the
decay products are neglected,
${\cal B}_{\tilde B}^{\not L}={\cal B}_{\tilde W}^{\not L}$.

The lepton number violating matrix elements, $V_{LH}$,  are of
${\cal O}(b_I/\tilde m^2)$ and ${\cal O}(m^2_{I0}/\tilde m^2)$,
where $\tilde m$ is the scale of the soft supersymmetry breaking
terms. From phenomenological constraints, such as neutrino masses
\cite{Chemtob:2004,Abada:2002}, we know that various lepton number
violating couplings must be suppressed by, at least, a factor of
${\cal O}[(m_\nu/m_Z)^{1/2}]$ \cite{Banks:1995}. Making the natural
assumption that R-parity is an approximate symmetry broken by small
parameters $\leq{\cal O}(10^{-4})$, we conclude that, very likely,
$|V_{LH}|\leq{\cal O}(10^{-4})$ and, consequently, \be {\cal
B}^{\not L}\leq{\cal O}(10^{-8}). \ee With fine-tuning, however,
this constraint can be avoided.

\section{The CP Asymmetry $\epsilon$}
The CP asymmetry generated from gaugino decays is defined as follows:
\begin{eqnarray}
  \epsilon_i \equiv \frac{\Gamma(\tilde \lambda_i \rightarrow
    \tilde LH^{\dag})+\Gamma(\tilde \lambda_i \rightarrow
    \tilde HL^{\dag})-\Gamma(\bar {\tilde {\lambda_i}} \rightarrow \overline {\tilde LH^{\dag}})
-\Gamma(\bar {\tilde {\lambda_i}} \rightarrow \overline {\tilde
HL^{\dag}})}
    {\Gamma(\tilde \lambda_i \rightarrow
   \tilde LH^{\dag})+\Gamma(\tilde \lambda_i \rightarrow
   \tilde HL^{\dag})+\Gamma(\bar{\tilde {\lambda_i}} \rightarrow \overline {\tilde LH^{\dag}})
   +\Gamma(\bar{\tilde {\lambda_i}} \rightarrow \overline {\tilde HL^{\dag}})}
    \end{eqnarray}
This asymmetry is calculated by evaluating the imaginary part of
the interference term between tree-level and one-loop diagrams.
Neglecting the masses of the decay products, we obtain an upper
bound on $\epsilon$ which serves also as a good estimate if there
is no strong phase space suppression.

For the CP asymmetry generated in $\tilde B$ decays, we obtain
\be
\epsilon_{\tilde B} \lesssim
    6\alpha_2\sin(2 \phi_g)\sqrt{y_W}\left[\frac{5}{2(1-y_W)}
  +1-(1+y_W) \ln\left(\frac{1+y_W}{y_W}\right)\right],
\ee
where $y_W={m_{\tilde W}^2}/{m_{\tilde B}^2}$. For $y_W
\in[0.5,1.5]$ we get $\epsilon_{\tilde B}\lesssim 1$ and
$\epsilon_{\tilde B}$ will
be estimated as such. The CP asymmetry generated in $\tilde W$ decays,
$\epsilon_{\tilde W}$, can be obtained straightforwardly from the
expression for $\epsilon_{\tilde B}$ by replacing
$\alpha_2\to\alpha_1$ and $y_W\to y_W^{-1}$.

Finally, we note that the process $\tilde W'_3 \leftrightarrow L^{\pm}W^{\mp}$
\cite{Ma:2000,Hambye:2000zs,Ma:2000wp} does not occur at tree-level
because at high temperatures, before the EWPT, the gauginos do not mix with
the other neutralinos. Related processes, such as $\tilde W'_3
\leftrightarrow L^{\pm}W^{\mp}H'_{u,d}$, can occur at higher order.

\section{Departure from Thermal Equilibrium: $n_{\tilde B,\tilde W}/n_\gamma$}
The out-of-equilibrium condition reads
\begin{eqnarray} \label{eq:dfe}
   \Gamma_{\tilde X} < H(T=m_{\tilde X}) = 1.66g_*^{1/2}(m_{\tilde
   X}^2/M_{pl}),
 \end{eqnarray}
where $H$ is the Hubble constant, and $\Gamma_{\tilde X}$ is the
decay rate of the gaugino. For the bino, the decay rate is given by
\begin{eqnarray}
   \Gamma_{\tilde B} \sim \frac{5\alpha_1m_{\tilde B}}{8} \left\{ \begin{array}{ll}
  m_{\tilde B}/T & T \gtrsim m_{\tilde B} \\
  1 & T\lesssim m_{\tilde B}
 \end{array} \right.
\end{eqnarray}
while for the wino, one has to substitute $\alpha_1\to\alpha_2$,
$m_{\tilde B} \to m_{\tilde W}$ and $8 \to 2$. Putting these
expressions into eq. (\ref{eq:dfe}), we find that the out of
equilibrium condition is far from being satisfied for gaugino masses
of order a few TeV or less. Hence the bino and wino decouple much
below the temperature $T\sim m_{\tilde X}$ that is required in order
to have sufficient overabundance. We conclude that, at the time of
departure from thermal equilibrium, $n_{\tilde B, \tilde
W}/n_\gamma$ is exponentially suppressed and is well below ${\cal
O}(1)$. We now make a more quantitative estimate of this ratio.

 Let us examine the
processes that wash-out the lepton asymmetry that is generated in
gaugino decays. These processes are inverse decays, annihilations
and $2 \leftrightarrow 2$ scattering processes mediated by
intermediate gauginos. We follow similar considerations made in
refs. \cite{Kolb:1990,Riotto:1998bt}. It turns out that the most
significant process is the inverse decays. (This result is in
accordance with the Boltzman equations analysis carried out in refs.
\cite{Hambye:2000zs,Ma:2000wp}.) Thus we estimate the freeze-out
temperature $T_f$ as the one where inverse decays (ID) become
ineffective, $\Gamma_{ID}(T_{f})=H(T_{f})$. We obtain

\be T_f\lesssim m_{\tilde X}/36. \ee The ratio $T_f/m_{\tilde X}$ is
rather insensitive to variation of $m_{\tilde X}$ in the range of
$10^2-10^5$ GeV.

We make the approximation that from this moment and on the gauginos
propagate freely in the universe until they decay. Since the
temperature at the time of decay, $T_d^{\tilde X}$ defined by
$H(T_d^{\tilde X})\sim\Gamma_{\tilde X}$, fulfills $T^{\tilde
X}_{d}\gg m_{\tilde X}>T_{f}$, then the decays will be extremely
rapid. We can thus safely assume that all remaining gauginos decay
instantly at $T_f$. This approximation neglects the small fraction
of the asymmetry generated prior to $T_{f}$, and the small fraction
of the asymmetry that is washed out by inverse decays and scattering
that rarely occur at $T \lesssim T_{f}$. The number density at the
time of freeze-out is then estimated as follows:
 \be \left.\frac{n_{\tilde X}}{n_{\gamma}}\right|_{T=T_f}=
\frac{2}{\frac{2
        \zeta(3)}{\pi^2}T_f^3} \left(\frac{m_{\tilde X}T_f}{2\pi}\right)^{\frac{3}{2}}
        e^{-{m_{\tilde X}}/{T_f}}.
\ee Given the smallness of $T_f/m_{\tilde X}$, we find a very strong
suppression factor: \be \left.\frac{n_{\tilde
X}}{n_{\gamma}}\right|_{T=T_f}\sim 10^{-14}. \ee
\section{Conclusions}
We have analyzed the lepton asymmetry induced by bino and wino decays
with bilinear R-parity violating couplings. We obtained the following
results:

1. Lepton number violation at temperatures above the EWPT requires
misalignment between the lepton-Higgsino and slepton-Higgs mass
matrices. The resulting branching ratio is expected to be small, \be
{\cal B}^{\not L}_{\tilde B,\tilde W}\lesssim 10^{-8}, \ee unless
fine-tuning is involved.

2. The CP asymmetry is suppressed by a loop factor ($\alpha_1$ or
$\alpha_2$), but this could be compensated by other factors of order
one: \be \epsilon_{\tilde B,\tilde W}\lesssim 1. \ee

3. The gauge interactions cause the gauginos to depart from
thermal equilibrium at very late time, when their number density
is strongly suppressed: \be n_{\tilde B,\tilde W}/s \sim10^{-16}.
\ee 
This strong suppression factor related
to the
 late departure from thermal equilibrium is effective also in
 models with trilinear R-parity violating couplings.
 Actually, this situation has a much more general validity:
 If a decaying particle is not a singlet of the Standard Model
 gauge group, it will depart from thermal equilibrium at a
 temperature well below its mass (unless $M\gtrsim10^{14}\ GeV$).
 Its number density at the time of freeze-out would be too small
 to produce the observed baryon asymmetry (see {\it e.g.} \cite{Hambye:2001eu}).

We can thus estimate that the baryon number so generated is very
small,  \be B\lesssim 10^{-25}(\mathcal{B}^{\not L}/10^{-8}), \ee
to be compared with the observed value, $B\sim10^{-10}$. These
results prove that this model cannot account for the observed
baryon asymmetry of the universe. From a cosmological point of
view this failure adds up to the fact that R-parity violating
interactions mean that the LSP cannot serve as a good dark matter
candidate. Hence, from a cosmologist point of view, R-parity
violating models are more problematic (and less predictive) than
R-parity conserving ones.

 \section*{Acknowledgements}
I thank Yosef Nir for useful discussions, and Sacha Davidson and
Marta Losada for comments on the manuscript.

\end{document}